# Impact of excitation energy on hot carrier properties in InGaAs MQW structure

Hamidreza Esmaielpour[1], Laurent Lombez[2], Maxime Giteau[3], Jean-François Guillemoles[1,4], and Daniel Suchet[1,4]

[1] *CNRS-Institute Photovoltaïque d'Ile de France (IPVF), UMR IPVF 9006, 91120 Palaiseau, France.*
[2] *Laboratoire de Physique et Chimie des Nano-objets (LPCNO-INSA), 31077 Toulouse, France.*
[3] *Research Center for Advanced Science and Technology (RCAST), The University of Tokyo, Tokyo 153-8904, Japan.*
[4] *CNRS-Ecole Polytechnique, UMR IPVF 9006, 91120 Palaiseau, France.*

*Abstract* — Hot carrier solar cells aim to overcome the theoretical limit of single-junction photovoltaic devices by suppressing the thermalization of hot carriers and extracting them through energy selective contacts. Designing efficient hot carrier absorbers requires further investigation on hot carrier properties in materials. Although the thermalization of hot carriers is responsible for a large portion of energy loss in solar cells, it is still one of the least understood phenomena in semiconductors. Here, the impact of excitation energy on the properties of photo-generated hot carriers in an InGaAs multi-quantum well (MQW) structure at various lattice temperatures and excitation powers is studied. Photoluminescence (PL) emission of the sample is detected by a hyperspectral luminescence imager, which creates spectrally and spatially resolved PL maps. The thermodynamic properties of hot carriers, such as temperature and quasi-Fermi level splitting, are carefully determined via applying full PL spectrum fitting, which solves the Fermi-Dirac integral and considers the band-filling effect in the nanostructured material. In addition, the impact of thermalized power density and carrier scattering with longitudinal optical phonons on the spectral linewidth broadening under two excitation energies is studied.

## I. Introduction

Designing hot carrier absorbers capable of inhibiting hot carrier relaxation through electron-phonon interactions is of significant importance in advanced photovoltaic devices. In conventional solar cells, a large portion of absorbed solar energy is dissipated via thermalization of photo-generated hot carriers.[1] By harnessing the excess kinetic energy of hot carriers and converting it to electricity, it is possible to improve the efficiency of single-junction solar cells to 67% under one sun illumination.[2] To realize the objectives of such high efficiency solar cells, known as hot carrier solar cells, it is first required to slow down the thermalization rate of hot carriers in the absorber, then to extract them quickly before their kinetic energy is dissipated in the form of heat.[3]

Several experimental approaches have been proposed to reduce the thermalization rate of hot carriers, such as increasing the solar concentration,[3] designing nanostructured materials,[4-6] light trapping techniques in thin absorbers,[7-8] phonon engineering,[9-11] and extracting hot carriers stored in upper satellite valleys.[12-13] While the evidence of non-equilibrium hot carrier distribution has been observed in such designs, there are still open questions regarding the relaxation mechanisms of these particles.

Research on the properties of hot carriers has been carried out using various characterization techniques, such as steady-state photoluminescence (PL), [14] time-resolved photoluminescence, [15] transient absorption, [16-17] and electrical spectroscopy, [18] which has enabled us to broaden our knowledge of these non-equilibrium particles. However, the thermodynamic properties of hot carriers measured by these characterization methods, such as the temperature, are rarely compared to each other, mainly due to difference in experimental conditions.

Relaxation dynamics of hot carriers excited to different allowed energy levels in the band structure provides information about their relaxation mechanisms and their interactions with phonons. At a such condition, it is also possible to evaluate the performance of a hot carrier absorber in inhibiting the thermalization of hot carriers. In this work, discussions on the influence of excitation energy on the properties of photo-generated hot carriers at various lattice temperatures and excitation powers are provided. In addition, a new approach for fitting the full PL and absorption spectra is presented, which reduces the degrees of freedom of the fitting parameters, leading to a simplified analysis and more accurate results of the carrier temperature and the quasi-Fermi level splitting.

## II. Determination of carrier temperature and quasi-Fermi level splitting

The properties of photo-generated hot carriers in the InGaAs MQW structure are investigated using a hyperspectral luminescence imager under various continuous wave excitation wavelengths and lattice temperatures. By applying various calibration methods, the absolute photon flux emitted by the sample is determined. [19]

The un-doped MQW structure is grown by molecular beam epitaxy (MBE) and consists of five $In_{0.53}Ga_{0.47}As$ wells with 5.5 nm thick and $In_{0.8}Ga_{0.2}As_{0.44}P_{0.56}$ barriers (10 nm). The active region (the QWs and barriers) of the sample is isolated by un-doped InP cladding layers to improve the localization of photo-generated carriers within the QWs. The description of the growth parameters is discussed elsewhere. [20]

PL spectroscopy provides information about the optical and thermodynamic properties of emitting particles in the system. The emission spectrum can be modeled by the Planck's law of radiation, which considers a quasi-equilibrium condition, as described by: [21-23]

$$I_{PL}(E) = \frac{2\pi\, A(E)\, (E)^2}{h^3 c^2}\left[exp\left(\frac{E-\Delta\mu}{k_B T}\right) - 1\right]^{-1}, \quad (1)$$

where "$I_{PL}$" is the PL intensity, "$A(E)$" the energy-dependent absorptivity, "$h$" the Planck constant, "$k_B$" the Boltzmann constant, "$c$" the speed of light, "$T$" the carrier temperature, and "$\Delta\mu$" the quasi-Fermi level splitting. By fitting the full PL spectrum, the absorptivity of the sample and the hot carrier properties can be determined. This fitting method considers the band-filling effect in the absorptivity term, which is important especially at high excitation powers. In addition, driven by solving the Fermi-Dirac integral,[24] this fitting method can be applied to PL spectra with a wide range of $\Delta\mu$ (below and above the bandgap energy). Moreover, applying the full spectrum fit is important for nanostructured materials whose absorption spectra have energy dependence above the bandgap.

Here, the fitting processor is as follows: first, the temperature of carriers is found from the high energy side of the PL spectrum. Next, the relative quasi-Fermi level splitting of the PL spectra under various excitation powers is found. This relationship imposes a constraint on the degrees of freedom of the fitting parameters. Finally, by applying the full spectrum fit, the absolute values of the quasi-Fermi level splitting and the absorptivity of each individual spectrum are determined considering the impact of the band-filling effect on the absorptivity of the MQW structure as a function of the excitation power.

It is possible to estimate the carrier temperature from a *spectrally calibrated PL spectrum* by applying a linear regression to its high energy side in the range of energies in which the absorptivity is constant. This analysis can be performed by taking the derivative of Equation 1 respect to the photon energy under the Boltzmann approximation $\left(e^{\frac{E-\Delta\mu}{k_B T}} \gg 1\right)$, which is:

$$\frac{d}{dE}[\ln(I_{PL}(E)) - 2\ln(E)] = \frac{d}{dE}[\ln(A(E))] - \frac{1}{k_B T}. \quad (2)$$

Therefore, by plotting the left side of Equation 2, it is possible to determine the carrier temperature from the energy ranges where the slope of the graph is zero, which is attributed to the constant absorptivity. In other words, the offset of the region with a zero slope value is inversely proportional to the temperature of carriers. The key advantage of this method is to determine the range of energies with constant absorptivity, required for the basic linear fitting method. In bulk semiconductors, the absorptivity is (almost) constant above the bandgap, however, for quantum-well (QW) structures, the absorptivity has a step-like behavior, but it is still constant in the energy ranges between discrete energy levels. [20]

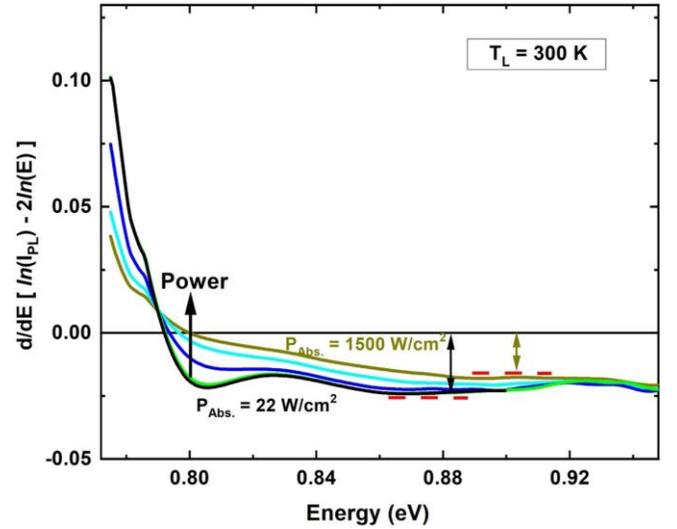

Figure 1. Derivative of the natural logarithm of the PL intensity emitted by InGaAs MQW at 300 K (minus $2Ln(E)$) with respect to the photon energy for various excitation powers. The red dashed lines indicate the range of energies where the slope of the curvature is zero.

In Figure 1, the carrier temperature associated with each spectrum is determined in the range of energies with zero slope values, shown by red dashed lines. It is observed that by increasing the excitation power, the range of energies in which the derivative of the curve becomes zero shifts to higher energy values, attributed to the band-filling mechanism. Without considering this effect and applying the linear fit to the same range of energies of the PL spectra emitted under various excitation powers, the extracted carrier temperatures become overestimated. Therefore, by finding the range of energies with constant absorptivity, it is possible to determine the temperature of carriers by applying the linear fitting method. At higher energies, as seen in Figure 1, a deviation from a straight line in the derivative of the absorptivity is observed, which is attributed to the low signal-to-noise ratio at energy values far from the peak of the PL spectrum. The second derivative of Equation 1 with respect to energy is performed, as shown in supplementary Figure S1, which indicates a zero value in the range of energies with constant absorptivity, determined from Equation 2. This result confirms that the offset values in Figure 1 is attributed to the inverse of the carrier temperature.

Finding the temperature of carriers from the derivative method requires the PL spectrum possesses a region with a constant absorptivity. However, if the absorption spectrum does not show an energy-independent behavior above the bandgap, this method cannot be applied. This can be the case in a QW structure with multiple discrete energy levels with low energy separation or in a sample with optical resonances (Fabry–Pérot oscillations) in its emission spectrum. To address such cases, a second method is proposed, which finds the carrier temperature directly from the absorption spectrum, as given by:

$$\ln(A(E)) = \ln(I_{PL}(E)) - 2\ln(E) + \frac{E}{k_B T} - \frac{\Delta\mu}{k_B T} + \ln\left(\frac{h^3 c^2}{2\pi}\right). \quad (3)$$

In Equation 3, the term $\left(\frac{\Delta\mu}{k_B T}\right)$ is constant at a given excitation power and it only changes the magnitude, but not the shape of the absorption spectrum. However, $\frac{E}{k_B T}$ in Equation 3, which is inversely proportional with the carrier temperature determines the slope of the absorptivity spectrum, especially at the high energy side. Therefore, the shape of the absorption spectrum can be determined by knowing the temperature of the emitting particles.

In this method, as shown in Figure S2(a) in the supplementary, one starts by plotting the absorption spectrum associated with the lowest excitation power (black), whose carrier temperature is equal to the lattice. Then, the next absorption spectrum is superimposed on the first one having the same slope on the high energy side by adjusting the temperature of carriers. The range of energies where the absorption spectra are to be superimposed is determined from Figure 1. Via this technique, it is possible to determine the carrier temperature associated with each individual spectrum. The legend of Figure S2(a) indicates the values found for each spectrum by applying this method. The comparison between the results determined by the two techniques is shown in Figure S2(b), which indicates good agreement between them.

The determination of the quasi-Fermi level splitting from the absolute calibrated PL spectrum, $I_a(E)$, can be performed by taking an integral from Equation 3, as described by:

$$\int_{E'}^{E''} Ln(I_a(E))dE$$
$$= \int_{E'}^{E''}\left[2Ln(E) + Ln(A_a(E)) - \frac{E}{k_B T_a} + \frac{\Delta\mu_a}{k_B T_a} + Ln\left(\frac{h^3 c^2}{2\pi}\right)\right]dE, \quad (4)$$

where $E'$ and $E''$ are the range of energies where the absorption spectrum is not influenced by the band-filling effect, as shown in Figure S3 in the supplementary. Integration over a wide energy range improves the accuracy of the results by mitigating fluctuations in the spectrum. The energy interval can also be selected where the absorption spectrum shows *energy-dependent* behavior. Thus, this method is not limited only for an absorption spectrum with a constant value above the bandgap. However, the absorptivity must be independent of the excitation power in the energy range of the interval.

By integrating another spectrum, $I_b(E)$, on the same energy range, $[E', E'']$, and by looking at the difference with, $I_a(E)$, the relation becomes:

$$\int_{E'}^{E''}\left[Ln(I_a(E)) - Ln(I_b(E))\right]dE$$
$$= -\int_{E'}^{E''}\left(\frac{1}{k_B T_a} - \frac{1}{k_B T_b}\right)E\,dE + \int_{E'}^{E''}\left(\frac{\Delta\mu_a}{k_B T_a} - \frac{\Delta\mu_b}{k_B T_b}\right)dE. \quad (5)$$

In Equation 5, the term due to the absorptivity is canceled, because the integral is taken from the range of energies with the same absorption values for both spectra, as shown in the highlighted region of the Figure S3(a) in the supplementary.

From Equation 5, $\Delta\mu_b$ can be found as a function of $\Delta\mu_a$, which is:

$$\Delta\mu_b = \frac{T_b}{T_a}\Delta\mu_a + \frac{k_B T_b}{(E'' - E')}\left([Intg(I_a(E)) - Intg(I_b(E))]_{E'}^{E''} +\right.$$
$$\left. + \frac{1}{2}\left(\frac{1}{k_B T_b} - \frac{1}{k_B T_a}\right)(E''^2 - E'^2)\right). \quad (6)$$

In Equation 6, "$Intg(I_a(E))$" represents the integrated PL intensity from $E'$ to $E''$, which can be determined directly from the absolutely calibrated PL spectrum. Equation 6 indicates that to find the absolute value of $\Delta\mu_b$, it is required to know the value for the reference spectrum ($\Delta\mu_a = \Delta\mu_{ref}$). Although Equation 6 does not directly determine the absolute value of $\Delta\mu$ for each individual PL spectrum, it reduces the degrees of freedom of the fitting variables by imposing a constraint, which will then lead to finding the absolute values of the absorptivity and $\Delta\mu$.

By applying Equation 6, it is possible to plot the relative absorptivity of multiple absorption spectra, as shown in Figure S4(a). Although the absolute value of the absorptivity has not yet been determined (it depends on $\Delta\mu$ based on the generalized Planck's law), the relative separation of the absorption spectra remains the same. To determine the absolute values of A(E), a full spectral fit is applied considering various optical transitions in the MQW structure, as given by: [25-26]

$$A(E) = [1 - R(E)] \cdot (1 - exp[-(\alpha_w d_w + \alpha_b d_b)]). \quad (7)$$

where "$R$" is reflectivity, "$\alpha$" the absorption coefficient, an "$d$" the absorber thickness. The indices "$w$" and "$b$" refer to the QW and the barrier, respectively. The absorption coefficient "$\alpha_{w0}$" of the potential well is determined by:[6, 23]

$$\alpha_{w0}(E) = a_x \cdot \exp\left[-\frac{(E-E_x)^2}{2\Gamma_x^2}\right] + \quad (8)$$

$$+ \sum_{i=1}^{n} a_i \cdot \frac{1}{1 + \exp\left(-\frac{E-E_i}{\Gamma_i}\right)} \cdot \frac{2}{1 + \exp\left(-2\pi\sqrt{\frac{R_y}{|E-E_i|}}\right)},$$

and the absorption coefficient "$\alpha_{b0}$" of the barrier:

$$\alpha_{b0}(E) = a_b \cdot \frac{1}{1+\exp\left(-\frac{E-E_b}{\Gamma_b}\right)}. \quad (9)$$

where "$a$" is the amplitude, "$E$" the energy of the optical transition, "$\Gamma$" the spectral linewidth broadening, and "$R_y$" the effective Rydberg energy of the material. The first term in Equation 8 is attributed to the excitonic transition and it has a large contribution at zero wavenumber (k = 0).[25] The second term in Equation 8 is due to band-to-band transitions from multiple discrete energy levels ($i = 1, ..., n$) in the MQW structure.

The band-filling effect in the QW structure is described by: [6, 27]

$$\frac{\alpha}{\alpha_0} = f_v^e - f_c^e$$

$$= \frac{\sinh\left(\frac{E - \Delta\mu}{2k_B T}\right)}{\cosh\left(\frac{E - \Delta\mu}{2k_B T}\right) + \cosh\left[\frac{m_h - m_e}{m_h + m_e} \cdot \frac{E - E_g}{2k_B T} - \frac{1}{2}\ln\left(\frac{m_h}{m_e}\right)\right]}.$$

(10)

where "$f_v^e$" and "$f_c^e$" are respectively the electronic distributions in the valence and the conduction bands. "$m_h$" and "$m_e$" are the effective mass of holes and electrons, respectively. Due to the band-filling effect, by increasing the excitation power, the number of states available in the conduction and valence bands decreases, resulting in a drop in absorptivity.[27] Therefore, the absorptivity of the sample becomes dependent on the thermodynamic properties (T and $\Delta\mu$) of the emitting particles, as indicated in Equation 10 through the band-filling effect.

The absorptivity of the MQW structure under various excitation powers is plotted in Figure S4(a) in the supplementary. The absorption spectra indicate a step-like behavior, especially at low excitation powers, attributed to the 2D density-of-states (DOS) of the MQW with two dominant discrete energy levels ($E_1 = 0.79$ eV and $E_2 = 0.83$ eV). Figure S4(a) shows that by increasing the excitation power, the absorptivity of the sample reduces (it becomes more transparent) at energy values close to the discrete energy levels. At very low excitation powers, an excitonic transition is observed at energy values close to the ground state transition, $E_1$, but not at the first excited state ($E_2$). Therefore, the drop in absorptivity at low photon energies, near $E_1$ is attributed to the combination of exciton dissociation and the band-filling effect. However, at energy values close to the first excited state, $E_2$, the drop in the absorption spectrum is attributed to the band-filling effect. Thus, by modeling the relative separation of the absorption spectra due to the band-filling effect, it is possible to estimate $\Delta\mu_{ref}$ responsible for such a separation.

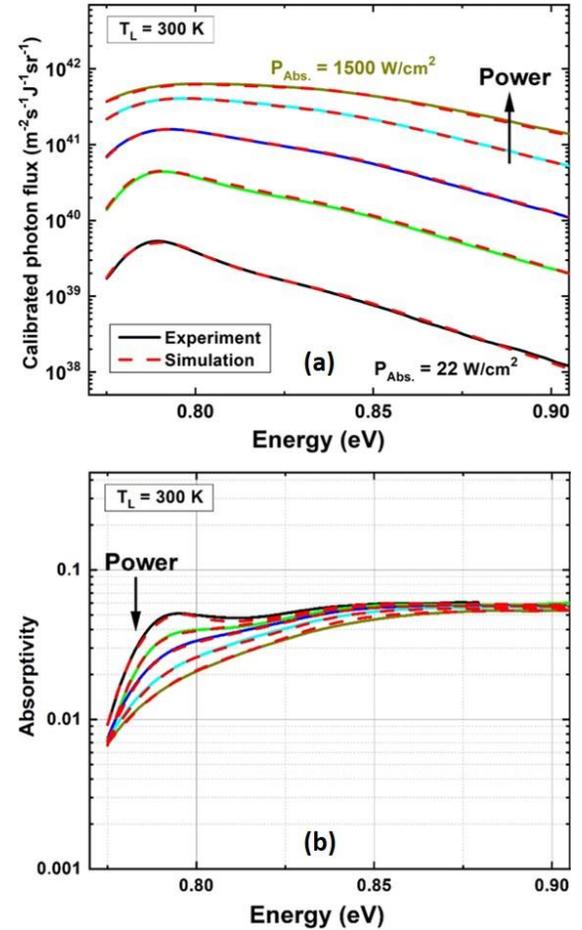

Figure 2. (a) Excitation power-dependent PL and (b) absorption spectra of the InGaAs MQW structure at 300 K. The solid and the dashed lines are respectively the experimental and the fitting results.

Figure S4 (b-d) indicates the simulation results for various $\Delta\mu_{ref}$ for the full fitting method. It is seen that only $\Delta\mu_{ref} = 0.54\ eV$, as shown in Figure S4(c), shows the best fitting results for the absorptivity of the MQW structure. Therefore, by finding $\Delta\mu_{ref}$, it is possible to determine the absolute values of the absorptivity and $\Delta\mu$ for all other spectra. Figure 2 shows the results of the full spectrum fit for (a) the PL spectra, and (b) the absorption spectra under various excitation powers at 300 K. A good agreement between the experimental and the simulation results is observed in both panels.

At very high excitation powers, when the quasi-Fermi level splitting is of the order of the bandgap or even larger ($\Delta\mu \geq E_g$), it is still possible to apply this method. However, first it is required to find the absorptivity of the sample at lower excitation powers, then, at higher powers the value of $\Delta\mu$ can be determined by considering the band-filling effect and by solving the Fermi-Dirac integral in generalized Planck's law.

## III. EXPERIMENTAL RESULTS AND DISCUSSIONS

The results of the carrier temperature difference ($\Delta T$: the difference between the carrier and the lattice temperatures) as a function of the absorbed power density, "$P_{Abs}$", at two lattice temperatures (300 K: green triangles and 230 K: red dots) and two excitation wavelengths (980 nm: solid symbols and 405 nm: open symbols) are plotted in Figure 3(a). The absorbed power density at each excitation wavelength is determined by the transfer matrix method.[28-29] The refractive indices of bulk semiconductors are applied for the analysis. The size of the laser spot is determined from its hyperspectral image at the full width at half maximum (FWHM) intensity.

Figure 3(a) shows that at 300 K upon increasing the absorbed power density, the hot carrier effect becomes stronger; i.e. $\Delta T$ increases, for both excitation wavelengths. However, the results indicate that the rise in $\Delta T$ under the higher excitation energy (405 nm) is larger than that of the lower energy photons (980 nm). A similar dependence of the excitation wavelength for the temperature of hot carriers at a given absorbed power density is observed at the lower lattice temperature (230 K), see Figure 3(a). Therefore, it is noted that the photo-generated hot carriers behave differently at a given absorbed power density under various excitation wavelengths.

To assess the performance of the sample in inhibiting the thermalization of hot carriers, it is required to consider the fraction of the absorbed power density above the band edge, i.e. the intra-band thermalized power density, "$P_{th}$", which is given by:[30]

$$P_{th} \approx \frac{E_{laser} - E_{gap}}{E_{laser}} P_{abs}, \qquad (11)$$

where "$E_{laser}$" and "$E_{gap}$" are the laser and the bandgap energies, respectively.

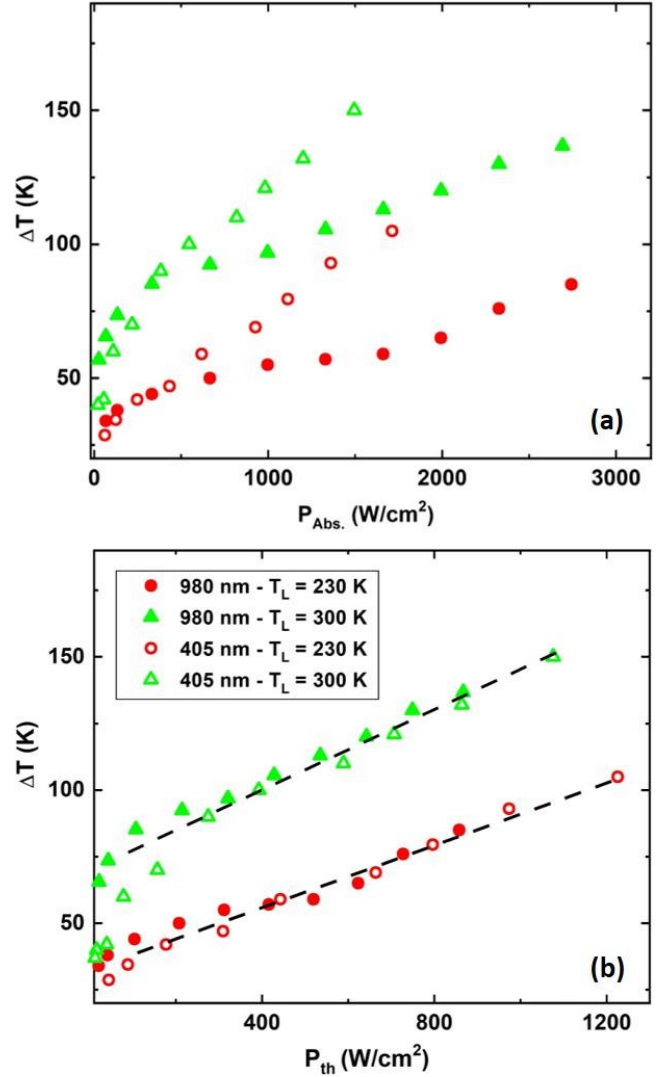

Figure 3. $\Delta T$ versus (a) the absorbed power density, and (b) the thermalized power density, at two lattice temperatures (230 K: red dots and 300 K: green triangles) and two excitation wavelengths (980 nm: solid symbols and 405 nm: open symbols). The black dashed lines show the gradient of the curve at two lattice temperatures.

Figure 3(b) shows the results of $\Delta T$ versus the thermalized power density applied for the same data set in panel (a). It is seen that by considering the thermalized power density, the dependence of $\Delta T$ on the excitation wavelength at 300 K is removed and the temperature of photo-generated hot carriers increases with the same rate under both excitation wavelengths. Similar behavior has also been observed in thin film GaAs absorbers of various thickness under two different excitation wavelengths.[30] In addition, it is seen that at the lower lattice temperature (230 K), the temperature dependence of the hot carriers with respect to the thermalized power density remains the same under two excitation wavelengths. The slope of the graph in Figure 3(b) is inversely proportional to the thermalization coefficient (Q) of the MQW structure, as described by:[30-31]

$$P_{th} = Q \cdot \Delta T \, \exp\left[-\frac{\hbar\omega_{LO}}{k_B T}\right], \quad (12)$$

where "$\hbar\omega_{LO}$" is the longitudinal optical (LO) phonon energy of the QW structure. This empirical coefficient estimates the strength of the thermalization rate of hot carriers through phonon-mediated relaxation pathways in the system. In Equation 12, the exponential term, which is related to the population of LO-phonons (from the Bose-Einstein distribution), is considered due to relatively large $\Delta T$ at the two lattice temperatures. The thermalization coefficients of the MQW structure at 300 K and 230 K are about 35 W K$^{-1}$cm$^{-2}$ and 65 W K$^{-1}$cm$^{-2}$, respectively.

The quasi-Fermi level splitting versus the thermalized power density is plotted for various lattice temperatures and excitation wavelengths in Figure 4(a). The two dashed horizontal lines indicate the position of the effective bandgap energies at 230 K (red) and 300 K (green). By increasing the lattice temperature, the effective bandgap of the MQW structure shifts to lower energy values due to the expansion of the crystal structure.[32] As shown in Figure 4(a), $\Delta\mu$ increases with the thermalized power density attributed to the increase in the density of photo-generated carriers at higher excitation powers. In addition, it is seen that by decreasing the lattice temperature, $\Delta\mu$ at a given thermalized power density becomes larger. This effect is attributed to the lower rate of non-radiative recombination and the reciprocal relationship between the carrier temperature and $\Delta\mu$ in the density of carriers under non-equilibrium conditions.[33-34]

Figure 4(a) also indicates that the rate of change of $\Delta\mu$ decreases at high thermalized power densities. To assess the origin of this effect, the relationship between the absorbed power density and the intensity of PL emission is studied, which allows to identify the type of dominant recombination mechanism in the system. [35] The results of the natural logarithm of the absorbed power density at the 405 nm excitation wavelength versus the natural logarithm of the integrated PL intensity at 300K and 230 K are plotted in Figure 4(b). Due to the similar behavior observed for the two excitation wavelengths, only the results of the 405 nm laser are plotted here. The relationship between the absorbed power density and the integrated PL intensity is described by: [35-36]

$$P_{Abs.} = A \, I_{PL}^{1/2} + B \, I_{PL} + C \, I_{PL}^{3/2}, \quad (13)$$

where "$A$", "$B$", and "$C$" are the coefficients attributed to Shockley-Read-Hall (SRH), radiative, and Auger recombination mechanisms, respectively. According to Equation 13, if the change of the absorbed power density with respect to the integrated PL intensity follows a power of "$1/2$", the dominant recombination mechanism is due to SRH, if it follows a power of "1", radiative, and "$3/2$", Auger recombination mechanism. The slopes of the linear fitting of $\ln(P_{Abs.})$ versus $\ln(I_{PL})$ are shown in the legends of Figure 4(b). The results indicate that at lower absorbed power densities, the dominant transition in the system is due to the radiative recombination (slopes are close to 1), and as the absorbed power density increases, the Auger recombination becomes more dominant (slopes become close to $3/2$). The strong contribution of Auger recombination, observed at the two lattice temperatures, explains the origin of the plateaus for $\Delta\mu$ at high thermalized power densities. In addition, the greater separation between the maximum $\Delta\mu$ and the bandgap energy, see the dashed lines in Figure 4(a), at the elevated lattice temperature is attributed to the combination of Auger heating and the greater hot carrier temperature at 300 K.

The dynamics of hot carriers is also reflected in the spectral linewidth broadening of their optical transitions in the system. The origin of linewidth broadening of a PL spectrum can be associated with the scattering of carriers with longitudinal optical (LO) phonons, acoustic phonons, ionized impurities, excitons, and charge screening effects.[37-39] At elevated lattice temperatures, the scattering of carriers with optical phonons in polar semiconductors (like group III-V) via the Fröhlich interaction is dominant.[37] At higher thermalized power densities, the population of non-equilibrium LO-phonons (or hot phonons) increases, which enhances the rate of carrier scattering with LO-phonons.[40] This effect can increase the linewidth broadening of hot photoluminescence spectrum at high thermalized power densities.

Figure 5 shows the change in the linewidth broadening at 300 K as a function of the thermalized power density under two excitation wavelengths. The figure shows the results of each individual optical transition in the MQW structure determined from the full spectral fit. Figure 5 indicates that by increasing the thermalized power density, the linewidth broadening of the excitonic transition ($\Gamma_x$: blue squares) increases. This effect is attributed to the power-dependent excitonic dissociation, which is consistent with the drop of the amplitude of this transition at high thermalized power densities, see Figure S7(b) in the supplementary.[39] The linewidth broadening for the first band-to-band transition ($\Gamma_1$: orange dots) shows a strong power-dependent behavior, unlike that of the second band-to-band transition ($\Gamma_2$: magenta triangles), which does not indicate a significant change as a function of power.

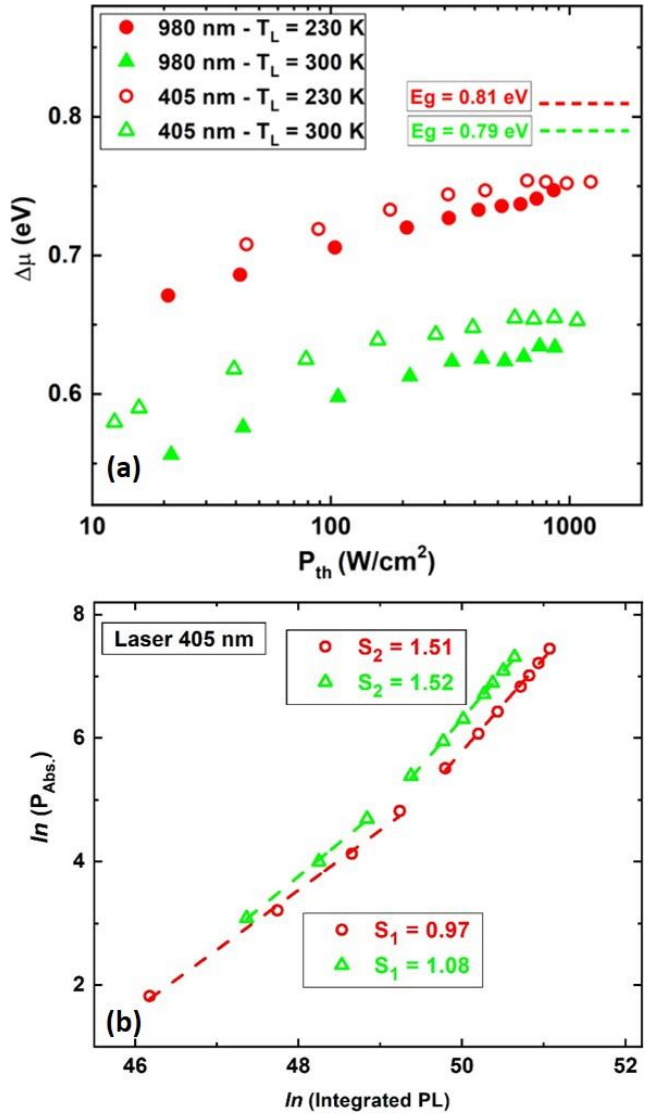

Figure 4. (a) $\Delta\mu$ of the photo-generated hot carriers as a function of the thermalized power density at various lattice temperatures (230 K: red dots and 300 K: green triangles) and excitation wavelengths (980 nm: solid symbols and 405 nm: open symbols). The dashed lines indicate the position of the effective bandgap energies at various lattice temperatures. (b) Natural logarithm of the absorbed power density versus the natural logarithm of the integrated PL intensity at 230 K (red) and 300 K (green) for the 405 nm excitation wavelength. The legends indicate the values of the slopes in the regions under low and high excitation powers.

To study the impact of carrier scattering by LO-phonons on the PL spectrum, the linewidth broadening of the first band-to-band transition emitted by photo-generated carriers at thermal equilibrium with the lattice ($T_C = T_L$) is determined. This low excitation power data is determined from PL spectra emitted by carriers far from the laser spot, which are detected by the hyperspectral luminescence imager. The indication of emission by thermalized carriers is shown in Figure S8; it is seen that at large distances from the laser spot in the hyperspectral image (greater than the diffusion length of hot carriers[41]), the shape of the PL spectra does not change at various distances, which indicates that the temperature of emitting particles remains constant in this region.[20] The broken x-axis at low powers in Figure 5 is attributed to a different relationship between the thermalized power density of the PL spectra emitted from regions far from the concentrated light with those at the center of the laser spot.

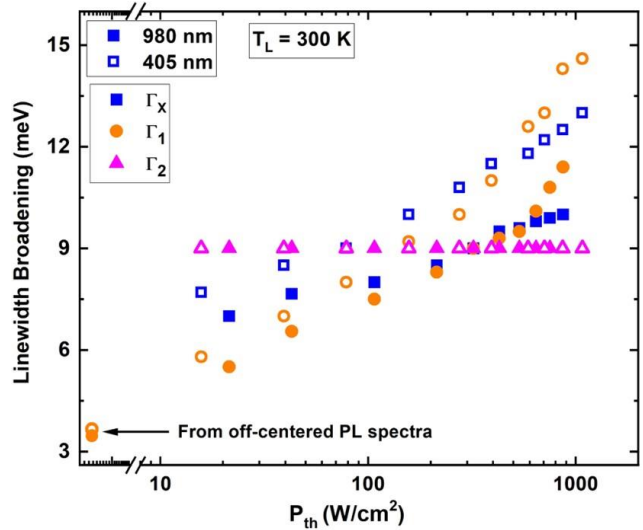

Figure 5. Spectral linewidth broadening at 300 K versus the thermalized power density for various optical transitions in the MQW structure: "$\Gamma_x$", excitonic (blue squares), "$\Gamma_1$", 1st band-to-band (orange dots), and "$\Gamma_2$", 2nd band-to-band (magenta triangles) transitions. The solid and open symbols are due to 980 nm and 405 nm excitation wavelengths, respectively. The spectral linewidth broadening attributed to the emission by carriers at thermal equilibrium with the lattice located at distances far from the laser spot is shown at low power region. A broken x-axis is applied to this low power data to indicate its different power relationship between the carriers emitted far from the laser spot and those at its center.

The relationship between the spectral linewidth broadening and the population of LO-phonons at elevated lattice temperatures is described as: [37]

$$\Gamma_{\text{High temperature}}(T) \approx \gamma_{\text{LO}} \cdot N_{\text{LO}}(T) = \frac{\gamma_{\text{LO}}}{\left[\exp\left(\frac{\hbar\omega_{\text{LO}}}{k_B T}\right) - 1\right]}, \quad (14)$$

where "$\gamma_{\text{LO}}$" is the Fröhlich coupling constant and "$N_{\text{LO}}$" LO-phonon population. At the thermal equilibrium condition, the temperature of LO-phonons is equal to the lattice ($T_{LO} = T_L$), therefore, the Fröhlich coupling constant under both excitation energies becomes $\sim 9.5\ meV$, which is in good agreement with the values reported for InGaAs MQW structures.[42] At higher thermalized power densities, where the effect of non-equilibrium hot carriers becomes evident, applying Equation 14 to find the spectral linewidth broadening will lead to values lower than the experimental results, see Figure S9 in the supplementary. This effect indicates that the determination of the spectral linewidth broadening under non-equilibrium conditions requires carrying out theoretical studies considering hot carrier scattering with hot phonons and charge screening effects, as seen by the wavelength-dependent behavior of the linewidth broadening.[43] Such a theoretical study provides insightful information on hot carrier dynamics in this MQW structure.

## IV. CONCLUSION

In conclusion, the effects of excitation wavelength on the properties of hot carriers in InGaAs MQW structure are studied. First, a developed model for the full PL spectrum fit is proposed, which facilitates the analysis and improves the accuracy of the fitting results. In this approach, it is discussed how to determine the unique values for the carrier temperature, the quasi-Fermi level splitting, and the absorptivity of the MQW structure from the PL spectrum by reducing the degrees of freedom of the fitting parameters.

The experimental results indicate that considering the absorbed power density, an excitation wavelength-dependent behavior is observed for $\Delta T$. However, the change of $\Delta T$ as a function of the thermalized power density is the same under both excitation wavelengths (405 nm and 980 nm), for different lattice temperatures. The results of $\Delta\mu$ as a function of the thermalized power density indicate that under low excitation powers, where the dominant recombination mechanism is attributed to radiative recombination, the gradient of $\Delta\mu$ is larger than when the Auger mechanism is dominant at high excitation powers. Finally, the impact of the thermalized power density on the spectral linewidth broadening is studied and the Fröhlich coupling constant of the MQW structure is determined.

ACKNOWLEDGEMENT

The research is carried out within the framework of program 6: PROOF (ANR-IEED-002-02) at IPVF. The authors would like to acknowledge the financial support of the French ANR project ICEMAN (No. ANR-19-CE05-0019). The author also would like to thank the Foton INSA Laboratory in Rennes, France, for providing the MQW structure.

# Supplementary Information

# Impact of excitation energy on hot carrier properties in InGaAs MQW structure

Hamidreza Esmaielpour[1], Laurent Lombez[1,2], Maxime Giteau[3], Jean-François Guillemoles[1,4], and Daniel Suchet[1,4]

[1] *CNRS-Institute Photovoltaïque d'Ile de France (IPVF), UMR IPVF 9006, 91120 Palaiseau, France.*

[2] *Laboratoire de Physique et Chimie des Nano-objets (LPCNO-INSA), 31077 Toulouse, France.*

[3] *Research Center for Advanced Science and Technology (RCAST), The University of Tokyo, Tokyo 153-8904, Japan.*

[4] *CNRS-Ecole Polytechnique, UMR IPVF 9006, 91120 Palaiseau, France.*

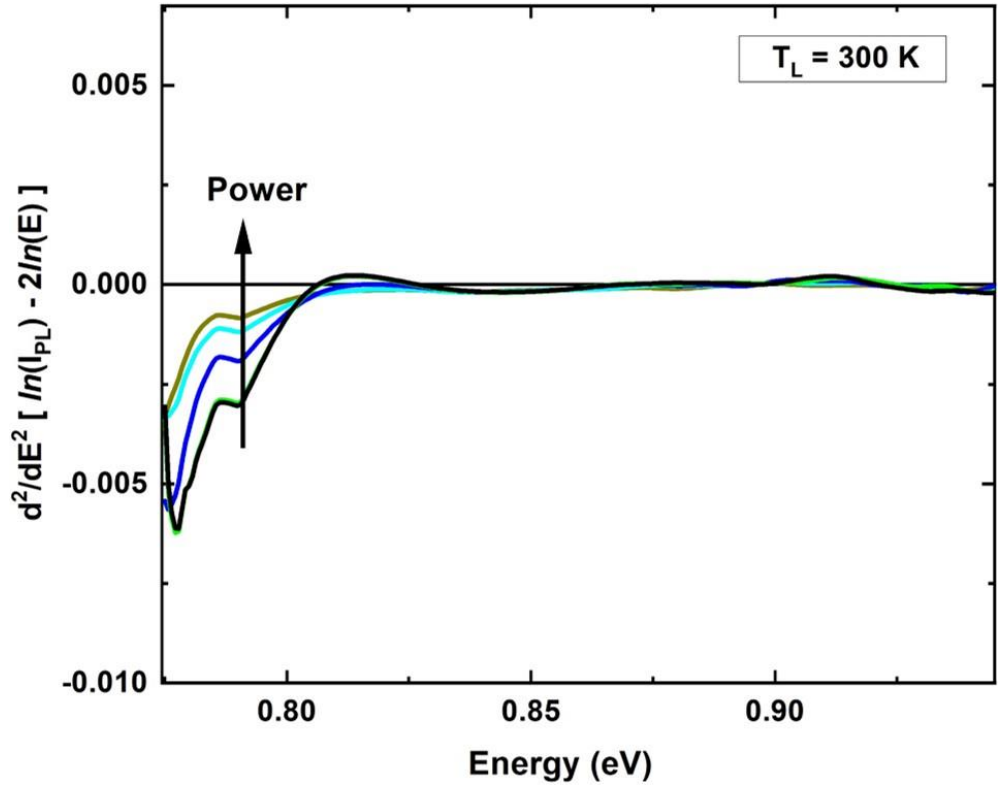

**Supplementary Figure S1:** Second derivative of $\ln\left(\frac{I_{PL}}{E^2}\right)$ versus the photon energy at 300 K and under various excitation powers. The figure confirms that at photon energies above (E > 0.85 eV), the absorption spectrum is independent of photon energy (or constant absorptivity) and excitation power (no band-filling), which are consistent with the results observed in the 1st derivative, as discussed in the manuscript.

$$\frac{d^2}{dE^2}[Ln(I_{PL}(E)) - 2\,Ln(E)] = \frac{d^2}{dE^2}\left[Ln\bigl(A(E)\bigr)\right].$$

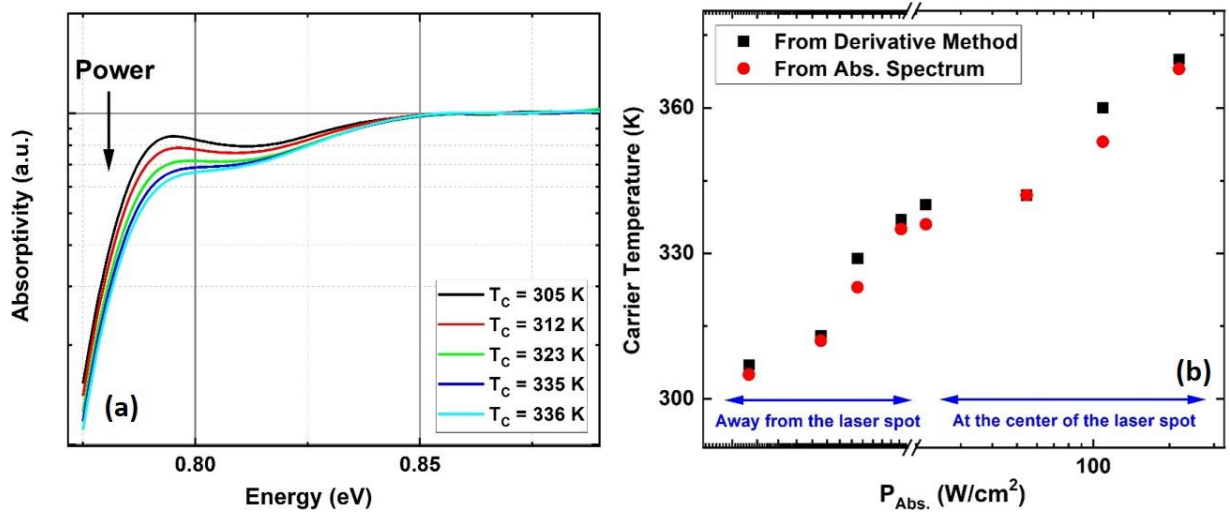

**Supplementary Figure S2:** (a) Absorption spectra of the MQW structure under low excitation power densities at room temperature. The black line shows the absorption spectrum of the sample at a condition, where emitting particles are at thermal equilibrium with the lattice ($T_C = T_L$). In Figure S8 of the supplementary it is discussed how to record the emission of carriers with a temperature equal to the temperature of the lattice. The temperature of carriers excited by higher power densities is determined by superimposing their absorption spectra on that of the lowest excitation power (black) in the range of energies in which no band-filling effect is present. This method can also be applied even if the slope of the absorption spectrum at the high energy side is not zero; in other words, the absorptivity can be energy-dependent. (b) The comparison between the carrier temperatures determined from the derivative method and the absorption spectra. A good agreement between the results of the two methods is observed. The x-axis shows a discontinuity because the first few temperatures under low excitation powers are found from the PL spectra emitted at positions far from the laser spot in the hyperspectral images.

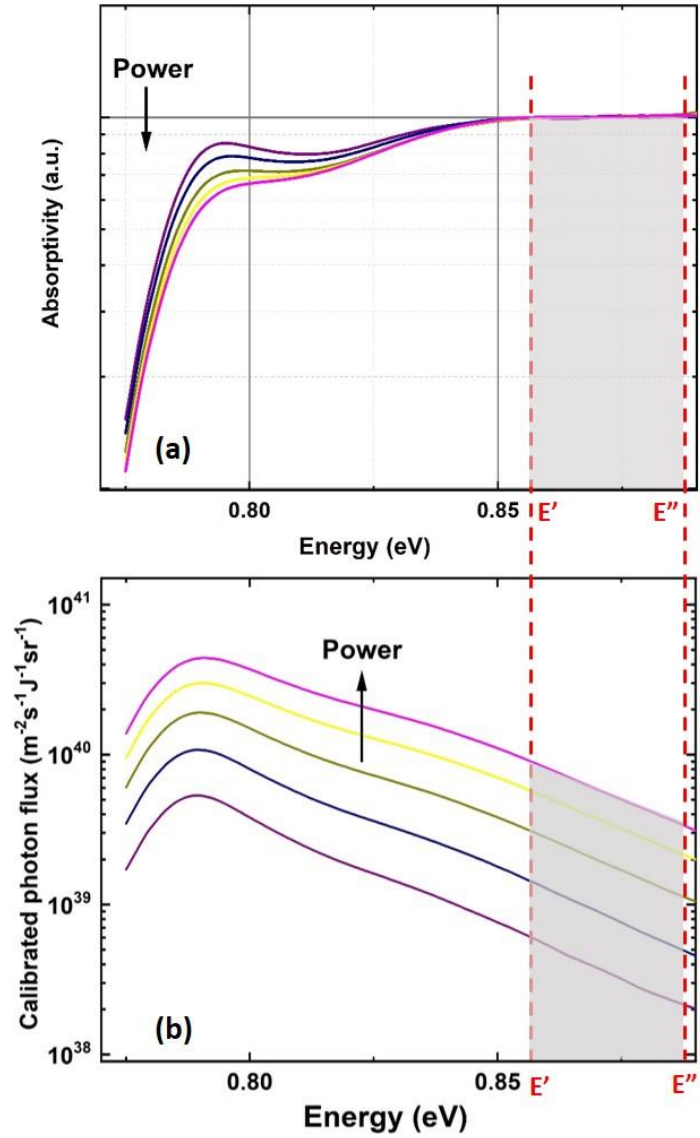

**Supplementary Figure S3:** Excitation power-dependent (a) absorption and (b) PL spectra at 300 K of the MQW structure. The shaded area indicates the range of energies with constant absorptivity where the analysis for the determination of the Quasi-Fermi level splitting is applied.

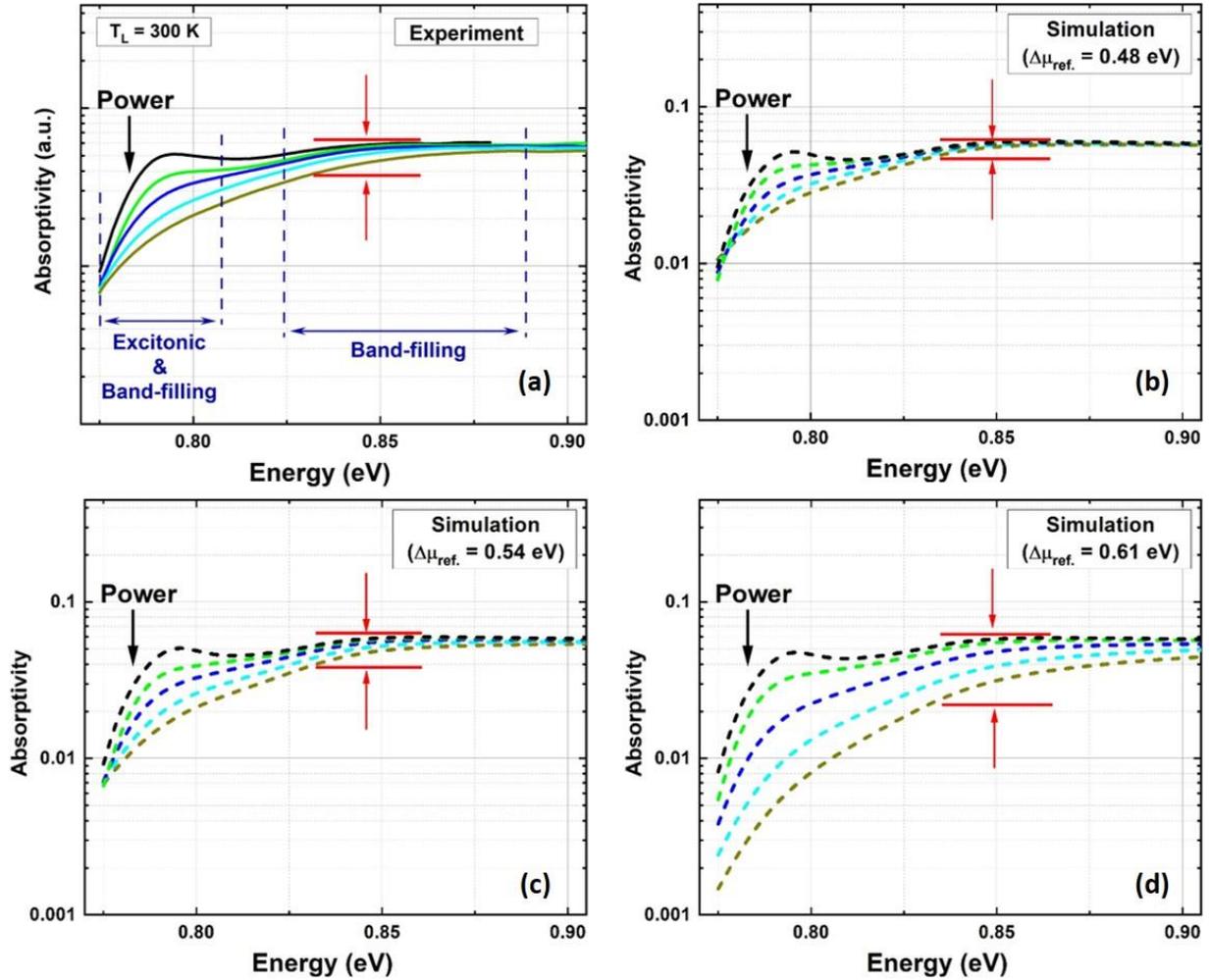

**Supplementary Figure S4:** (a) Excitation power-dependent absorption spectra of the MQW structure at 300 K. The y-axis in panel (a) has an arbitrary unit; to find the absorptivity of the sample from the generalized Planck law, it is required to determine the absolute value of $\Delta\mu$ of each individual spectrum. At low photon energies (E < 0.82 eV), the relative change in the absorption spectra at various excitation powers is due to the combination of the exciton dissociation and the band-filling effect. However, at the energy ranges around the second band-to-band transition (0.83 eV < E < 0.88 eV) in the MQW, as shown by the red solid lines in panel (a), no excitonic transition (or peak) is present and the relative change at various powers is attributed to the band-filling effect. To determine the absolute value of $\Delta\mu$ for each spectrum, a numerical simulation is carried out considering various optical transitions (one excitonic, and two band-to-band) in the sample, as shown in panels (b), (c), and (d). It is seen that by considering $\Delta\mu_{ref.} = 0.54$ eV, the spectral shape and the relative separation of the absorption spectra at various excitation powers are recovered.

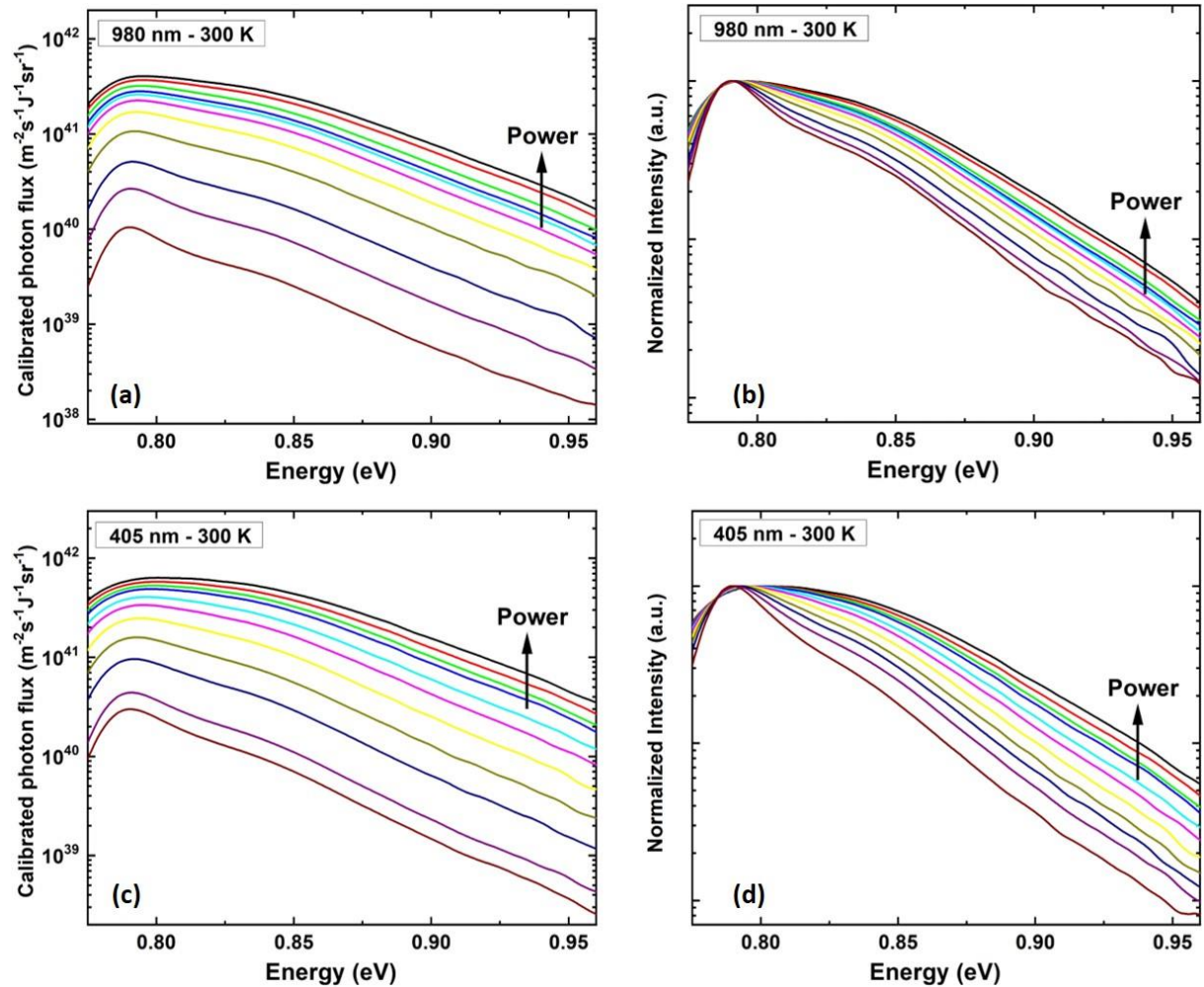

**Supplementary Figure S5:** Excitation power-dependent PL spectra at 300 K under (a) 980 nm and (c) 405 nm excitation wavelengths. The normalized PL spectra at 300 K are plotted in (b) and (d) for the excitation wavelengths of 980 nm and 405 nm, respectively.

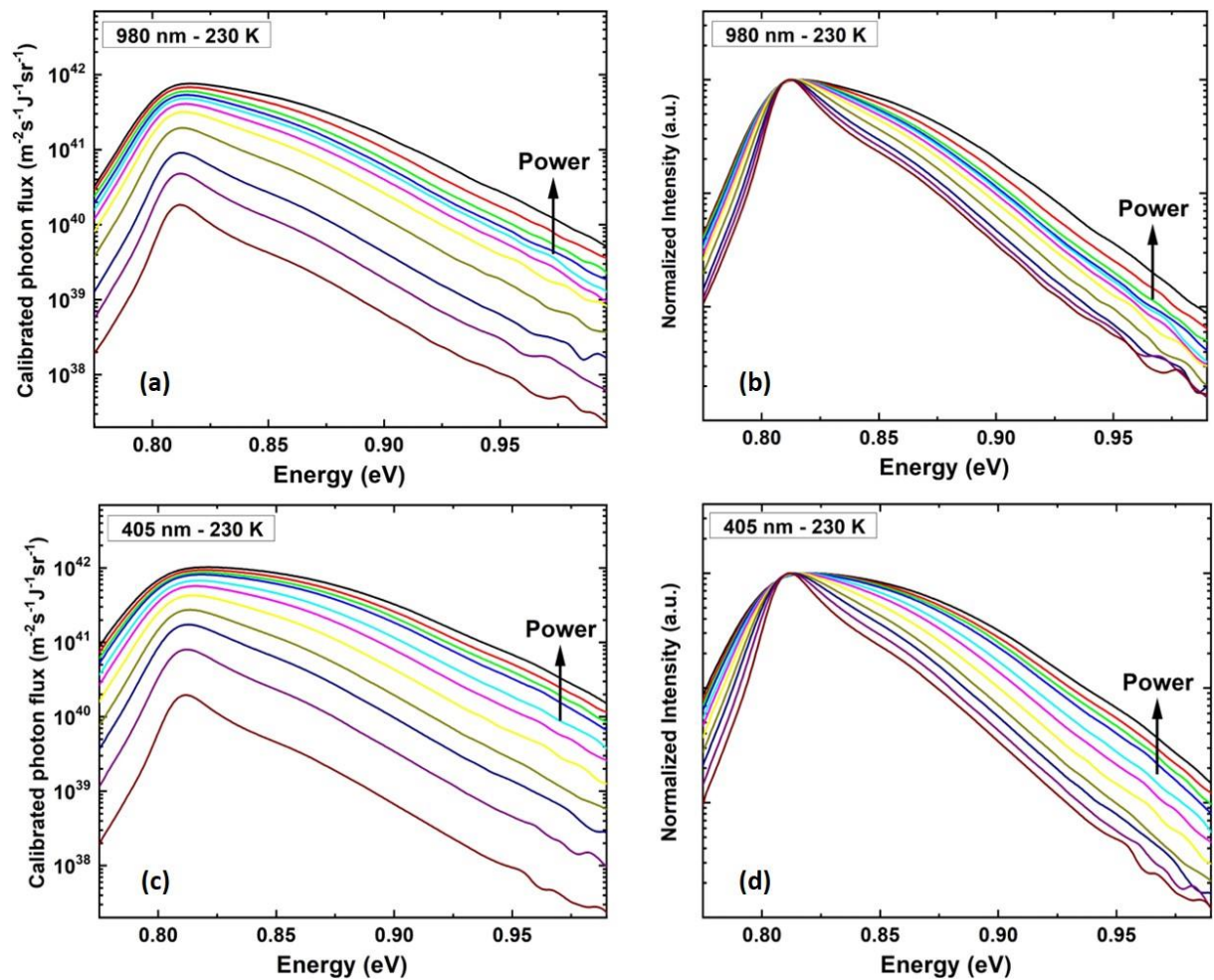

**Supplementary Figure S6:** Excitation power-dependent PL spectra at 230 K under (a) 980 nm and (c) 405 nm excitation wavelengths. The normalized PL spectra at 230 K are plotted in (b) and (d) for the 980 nm and 405 nm excitation wavelengths, respectively.

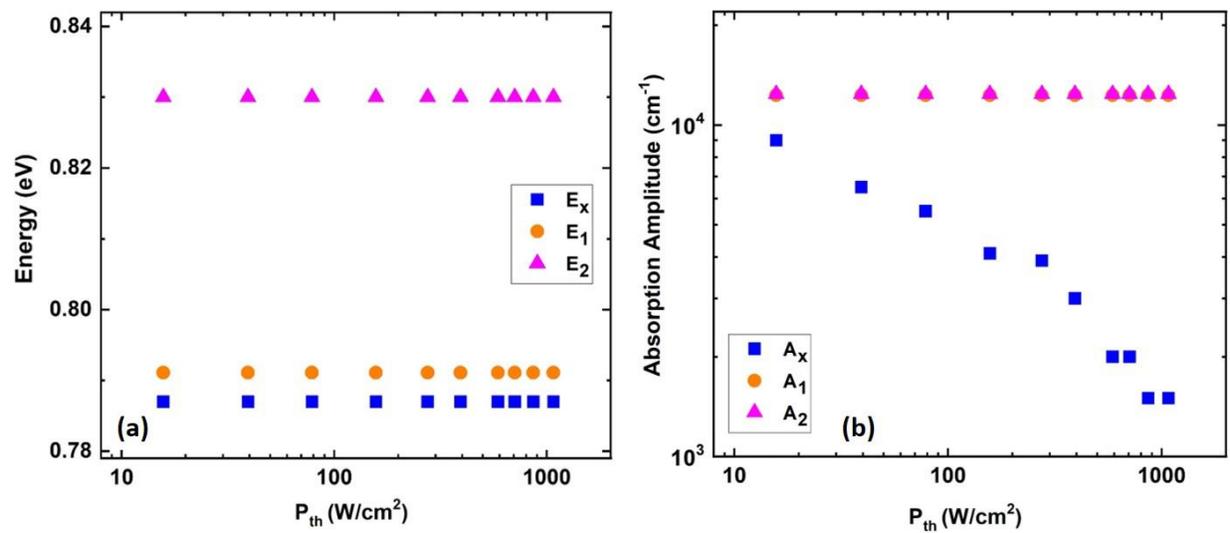

**Supplementary Figure S7:** (a) Energy and (b) absorption amplitude of the MQW structure associated with each individual optical transition.

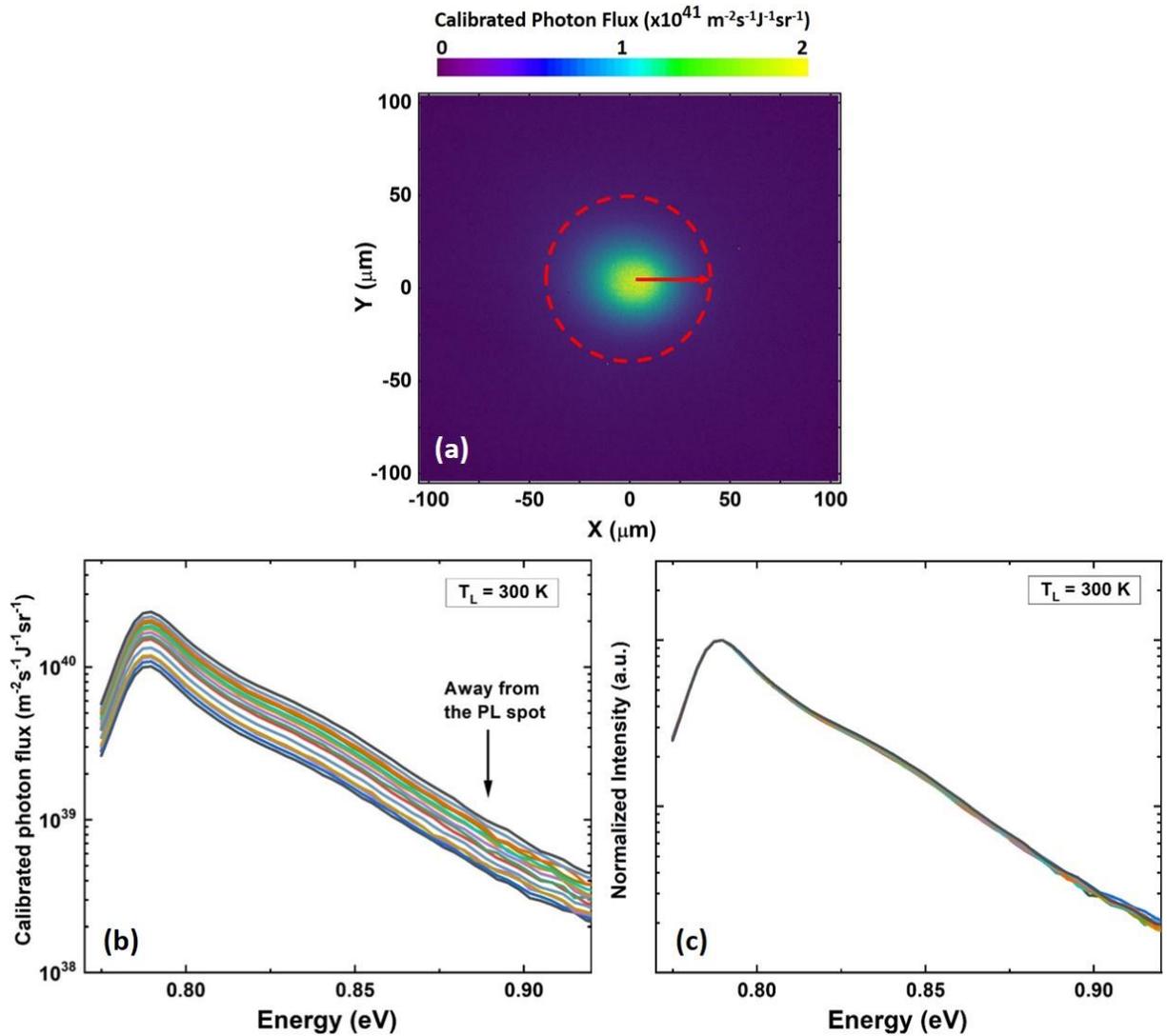

**Supplementary Figure S8:** (a) 2D-PL map of the InGaAs MQW emission under the lowest excitation power detected by the hyperspectral luminescence imager. The hot spot is the region, where the laser is concentrated. The red dashed line shows the area far from the laser spot ($> 40\ \mu m$), where carriers are in thermal equilibrium with the lattice. (b) PL spectra at various distances from the PL (or laser) spot from the region indicated by the red dashed line. (c) Normalized PL spectra of the same data set shown in panel (b). It is seen that at large distances from the laser spot, the shape of the PL spectra remains the same and only their intensity changes. This effect indicates that at such distances, the emitting particles are in thermal equilibrium with the lattice ($T_C = T_L$). These PL spectra are applied to determine the smallest linewidth broadening of each optical transition in the system.

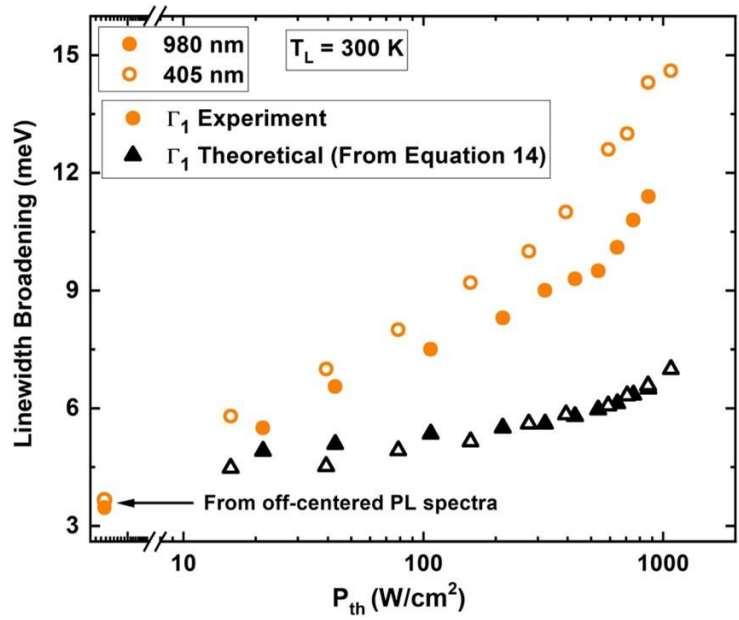

**Supplementary Figure S9:** Comparison between the experimental (orange dots) and the theoretical (black triangles) results of the linewidth broadening for the 1st band-to-band transition. It is seen that by considering only the contribution of hot phonons in the linewidth broadening determined by Equation 14, the calculated values are smaller than the experimental results.